**Main Manuscript for**

# Measurement of Nanoscale Interfacial Contact Area Using Vibrational Spectroscopy

Utkarsh Patil[1†], Stephen Merriman[1†], Shubhendu Kumar[1†], Ali Dhinojwala[1*]

[1]School of Polymer Science and Polymer Engineering, University of Akron, Akron, OH 44325 USA.

*Corresponding author, **Email:** ali4@uakron.edu

†These authors contributed equally to this work.

**Author Contributions:** U.P., S.M., S.K., A.D. designed research; U.P., S.M., S.K. performed experiments. U.P., S.M., S.K. performed data analysis; A.D. acquired funding and provided supervision; U.P., S.M., S.K., A.D. wrote and revised the manuscript.

**Competing Interest Statement:** The authors declare no competing interests.

**Classification:** Physical Sciences, Applied Physical Sciences

**This PDF file includes:**

> Main Text
> Figures 1 to 6
> Table 1

## Abstract

Understanding the extent of contact between objects is crucial for surface-driven phenomena in the natural world. While interactions are known to occur at subnanometer-length scales, the techniques needed to measure such small-length scales are lacking. We combined two complementary techniques of attenuated total reflectance infrared (ATR-IR) spectroscopy coupled with an imaging detector and sum frequency generation (SFG) spectroscopy to characterize the real contact area between surfaces. Imaging ATR-IR provides the distribution of airgap separation between the surfaces at larger than molecular length scales as well as an estimation of the real contact area. Additionally, SFG directly probes the intermolecular interactions occurring at molecular scales. Through the combination of these techniques, we show how contact is achieved for both thin, rigid/glassy polymer films on top of soft elastomers and rigid/glassy samples. Interestingly, we also show that the orientation of the molecules in contact is a critically important factor that influences interfacial interactions rather than just the number of molecular sites in contact.

## Main Text

### Introduction

Measuring the degree of contact between two surfaces has been a fundamental problem in many fields, where physical properties such as adhesion, friction, or conductivity are a function of the real contact area. In earlier studies, Amontons and Coulomb made puzzling observations that the friction force between sliding objects depends mostly on the normal load rather than the observed area of contact[1, 2]. Later, Bowden and Tabor explained these observations by noting that the actual contact area is important and, for rough surfaces, the contact area is a tiny fraction of the macroscopically observable contact area[3]. This "real area of contact" is what increases with normal load, independent of the apparent contact area. Thus, the interfacial properties of a system would be determined by the extent of direct contact between surfaces, which is known as *conformability*.

Fundamentally, the definition of the real contact area (molecular conformability) is ambiguous and depends on the physical property of interest. For example, when the distance between two opposing surfaces is around 0.16 nm, the van der Waals interactions are maximum based on the Hamaker equation (4). Since it is these interactions that govern processes like adhesion or friction, sites should be within this distance to be considered in complete contact. Further, separation distances of 1–3 nm become increasingly important for the case of electron tunneling effects, wherein electron transfer occurs between surfaces with relevance to redox reactions in biological proteins[5] and electrical conductivity for metallic systems[6]. Finally, the heat conductivity for rough surfaces is sensitive to the submicron airgap separations present at the interface[7]·[8]. Thus, an understanding of contact conformability at nano-length scales becomes highly relevant for understanding the magnitude of the contact properties at the interface. However, it has been challenging to precisely characterize molecular conformality down to the nanoscale using various available optical[9–16], fluorescent[17–20], or electrical[21, 22] methods due to limitations of spatial resolution, light scattering from rough surfaces, or lack of molecular-scale sensitivity. Consequently, experimental evidence of the molecular conformality achieved between surfaces remains an unresolved issue.

Here, we propose the use of two vibrational spectroscopic techniques—namely, attenuated total reflectance infrared (ATR-IR) and sum frequency generation (SFG) spectroscopy—to probe contact from separations at subnanometer to micron scales. In ATR-IR, similar to previous methods based on total internal reflection[14, 16], measured intensity is fundamentally related to the number of contacts or the airgap separation between the ATR crystal and a contacting material. However, since we record the intensity of a resonant absorption at the interface rather than the raw reflected light intensity, scattering effects can be readily accounted for[23]; thus, uncertainties for very rough surfaces or air gaps at small length scales may be reduced. We further apply a two-dimensional imaging detector as a means of visualizing the nature of the contact made in three dimensions.

In addition, SFG represents an interface-specific technique where a signal is only generated in non-centrosymmetric media (e.g., interfaces) with molecular-scale sensitivity[24]. Recently, SFG has been used to measure the interaction energy for liquids[25] and to predict the work of adhesion for solids in contact[26]·[27]. This ability derives from the shift of the vibrational peak of surface hydroxyl groups on a sapphire crystal after mechanical contact. The greater the peak shift, the stronger this acid–base interaction energy. We postulate that this interaction energy for contacting solids of a given material should also be fundamentally related to the number of molecules in contact, or the *molecular contact area*. Furthermore, since SFG is probing contact through interactions that occur at sites of contact, the technique directly assesses the property of molecular conformability by its very definition and at the smallest length scales. Since the acid–base interactions are short-range interactions, the detection of shift is incapable of providing details beyond those subnanometer points of contact. Through a combination of both ATR-IR and SFG spectroscopy, the length scales spanning molecular contact to submicron-scale surface separations can readily be bridged to fully characterize the molecular conformability of the contact interface.

In our experiments, we chose surfaces consisting of poly(methyl methacrylate) (PMMA) of varying effective modulus or roughness by either preparing thin bilayer PMMA films of varying thickness on soft elastomeric substrates to probe the effect of modulus or by preparing PMMA sheets treated with sandpaper of varying grit size to probe the effect of roughness. For soft elastic contacts, model-based predictions indicate that the surfaces should conform to achieve complete conformal molecular contact at even the smallest length scales, even though it remains a challenge to obtain direct experimental evidence of this (28, 29). For hard, rough contacts, previous experiments show that the surfaces should not be able to conform and that only a small amount of molecular contact is made(16, 18). However, the molecular contact for hard surfaces that are very smooth (root mean square average (RMS) < ~20 nm) has been difficult to characterize thus far with the resolution of existing methods. By applying imaging with ATR-IR and SFG to these chosen systems spanning modulus (soft to hard) and roughness (smooth to rough) for a given fixed material (PMMA), we aim to systematically investigate quantitatively the molecular contact made using our proposed techniques for the very first time.

## Results

### ATR-IR Imaging of Contacts

First, we employed ATR-IR spectroscopy as a tool to quantify and visualize contacts at the submicron-length scale. The samples chosen for analysis, alongside their respective modulus and roughness ($h_{rms}$) values, are provided in **Table 1**. More detailed roughness profiles are presented in power spectral density plots in **Figure S1**. The two bilayer samples of either a 10-nm or a 200-nm PMMA rigid film supported on a soft polydimethylsiloxane (PDMS) substrate and the smooth PMMA "hardsheet" (a thick sheet of pure PMMA molded against a smooth film) represent smooth samples with varying effective moduli. The three PMMA hardsheet samples that have been treated using sandpapers with different grits represent rigid samples with varying roughness, where papers with smaller grit values produce surfaces with higher roughness. In all cases, the surface chemistry in contact is that of the PMMA. A diagram of the experiment is provided in **Figure 1**, where a germanium hemispherical ATR crystal is brought into contact with a substrate at a certain indentation depth, $d$ (*i.e.*, a certain pressure). A two-dimensional focal plane array (FPA) detector is used to collect Fourier transform infrared (FTIR) spectra in a 64-pixel × 64-pixel grid, where each pixel corresponds to an independent absorption spectrum. Examples of such spectra are provided in **Figure S2**.

Images for each of the different substrate contacts at different applied pressures are shown in **Figure 2**. For the case of the 10-nm bilayer, the image was constructed using the 1260 cm$^{-1}$ peak assigned to the Si-CH$_3$ bond of PDMS(30). For all other samples, the peak at 1725 cm$^{-1}$ assigned to the C=O bond of PMMA(31) was used for the analysis. The peak at 1260 cm$^{-1}$ was chosen for the 10-nm bilayer because the intensity of the peak at 1725 cm$^{-1}$ was weak. Peak areas were determined by subtracting a baseline to remove effects such as scattering(23) and then integrating between the pre-defined bounds provided in **Table S1**. Pixels of the image with a greater peak area represent pixels where the surfaces are closer together on average (*i.e.,* with a smaller average airgap thickness) within the pixel size of 1.1 μm × 1.1 μm. For each case, the leftmost image represents the initial observable point of contact with increasing pressure toward the right. Since the two bilayer samples possess a much lower effective modulus compared to the hardsheets (**Table 1**), the first measurable instance of contact completely fills the viewable 70 μm × 70 μm window of the detector. For the smooth hardsheet, a circular contact spot is clearly visible and grows with increasing applied pressure. Within the contact spot, intensity appears largely homogeneous and similar to the bilayers. For the sandpaper-roughened hardsheets, the introduced macroscopic roughness serves to obscure the boundaries of the apparent area of contact, and we observe a highly heterogeneous contact.

To quantitatively analyze the different contacts, we directly calculated the airgap separations from peak areas from the spectral images. The procedure for this calculation is provided in the Supplementary Information. Results are plotted in **Figure 3 a,b** for each sample type as an average airgap thickness within the entire image frame. The horizontal axis is given as indentation depth to facilitate direct comparisons between samples of different moduli, representing a range of 0–30

MPa for hardsheets and a range of 0–10 kPa for soft bilayers under a pressure that is based on Hertzian mechanics. In **Figure 2**, we observe increasing macroscopic heterogeneity in contact with increasing roughness from the smooth hardsheet towards the 30-grit hardsheet. For these samples, this macro-scale roughness is reflected by the large air gaps (*i.e.*, gaps greater than 600 nm) on initial contact that compress down with increasing pressure (**Figure 3a**). Of note, the 30-grit and 1500-grit hardsheets follow distinctly different trends for shape, with the air gap compressing to 19% its initial value for the 1500-grit hardsheet as compared to only 55% of the initial value for the 30-grit samples. This can be rationalized from the contact images shown in **Figure 2** for the two samples. For the 1500-grit hardsheet, initial contact is made at only a few sparse points. Upon increasing pressure, these contacts are compressed; however, at the same time, the new points of contact being made are rapidly decreasing the average airgap thickness. In the case of the 30-grit hardsheet, initial contact is made along a single ridge within these length scales; any further increase in pressure will primarily deform the material just around this ridge, while the remainder of the space remains out of contact. Finally, the smooth hardsheet with the lowest roughness shows non-significant air gaps as compared to the other roughened sheets, owing to its small asperity heights at the surface as compared to the microscopic scale (**Table 1** and **Figure S1**).

Furthermore, considering the regime of smooth samples ($h_{rms}$ < 10 nm), we study the effect of modulus. The moduli for the samples, including the PMMA–PDMS bilayers and smooth PMMA hardsheet, are plotted in **Figure 3b**. As described in the Supplementary Information, an air gap of zero was defined for these samples based on the most intense region for each sample at the maximum applied pressure, making the reported values effectively relative to the lowest air gap achievable at the largest pressure. The scale of these air gaps is much less than for the roughened samples—no more than about 30 nm. Moreover, we observe no change in airgap thickness with increasing pressure for either of the soft bilayer samples, resulting in a flat line around air gaps of 7–10 nm. This suggests, within the sensitivity of this technique, that the surfaces achieve conformal contact, since pressure increases are incapable of reducing the air gap. The fact that the values are non-zero is indicative of the sensitivity for which air gaps can be determined in absolute terms, which appears to be smaller than around 10 nm. Although air gaps at smaller scale could exist, we are unable to measure them. A rather interesting observation about the smooth hardsheet, which has comparable $h_{rms}$ to that of the bilayers, is that it displays a larger air gap that decreases with pressure. This pressure-dependent behavior, which is resolved in increments of less than 5 nm (even smaller than the absolute sensitivity of the technique), suggests a non-conformal contact. Even though the roughness is similar, the large modulus compared to that of the bilayers prevents the asperities from fully conforming. Instead, the smooth surface provides for the length scales of the out-of-contact regions to be comparatively small, which this technique has the sensitivity to resolve. In **Figure 3c**, three-dimensional airgap maps help visualize the differences between the hard and soft samples. The rough sample has a larger-scale random height profile, while the smooth samples exhibit smaller-scale gaps. The germanium crystal itself is not perfectly smooth and possesses a surface asperity that becomes influential on the air gap only on the smooth samples, appearing as a slightly heightened central region that is consistent from sample to sample. For the smooth hardsheet, this region is much larger in scale than the bilayers and compresses with pressure, confirming it is a real air gap and highlighting the reduced compliance. For the bilayers, this region is small in height and shows no pressure dependence. Thus, we believe the calculated height is not due to a real air gap but rather to slight local changes in the incident angle of the light on the crystal, which can become significant when the air gaps are of such low values, and this is one factor contributing to our observed sensitivity.

Finally, we utilize spectral information to calculate a real contact area in terms of a percentage of the total area of the image frame—following the approach of Rubinstein, Cohen, and Fineberg(12, 16)—and extend it to FTIR absorbance. This is accomplished by utilizing the expectation that every point of contact, even those that are made at scales smaller than the 1.1 μm that we resolve, contributes a peak area intensity corresponding to that for an air gap of zero (**Figure S5**). Provided the points not in contact are outside of the evanescent wave decay length of around 190–200 nm, then only the points in contact contribute peak area intensity—and this intensity will be linearly related to the real contact area as $A_{real}(\%) = \frac{I}{I_{max}}$, where $I$ is the observed peak area intensity and $I_{max}$ is the maximum achievable intensity at complete contact. For out-of-contact air gaps less than 190–200 nm (such as for smoother samples), this calculation will overestimate the real contact area, as the evanescent wave spanning the small gap will contribute additional measured intensity. As the $h_{rms}$ increases, we expect this calculation to converge toward the true value.

The results of this analysis are plotted in **Figure 4a-c** for the high-modulus hardsheets of varying roughness, where we expect non-conformal contacts. The approximately linear relationships agree with those reported from similar systems,(16) and the magnitude of the area increases as the surfaces become smoother. The area achieved at an applied pressure of 25 MPa is compared between systems in **Figure 4d**. Increasing the surface roughness logarithmically decreases the real contact area, and very low percentages of the nominal area achieve contact for the very rough 30-grit sample. While these area percentages may be overestimated, especially as the sample surfaces become smoother, the differences between samples can also be determined by comparing their average airgap thicknesses, which are directly measured without the same assumptions. When comparing the average airgap thickness values for the samples at a pressure of 25 MPa (**Figure 4e**), we again observe the airgap thickness increasing exponentially with RMS roughness.

**Molecular contact area characterization using SFG.**

In addition to ATR-IR, we utilized SFG spectroscopy to characterize the contact made for each sample at molecular length scales, in this case using a sapphire crystal as the opposing surface. While smooth surfaces such as those in the bilayers may show complete contact at larger scales (*i.e.*, at scales probed by ATR-IR), further magnification to lower length scales (those of individual molecules) show partial contact between the two opposing surfaces(28) (**Figure 5c**). This molecular-level contact is susceptible to changes in local conformation of the molecules, the number of interacting species, and the types of interactions such as van der Waals interactions, acid–base interactions (a subset of which is hydrogen bonding), and electrostatic interactions. Thus, through direct probing of the molecular interactions between the contacting surfaces, surface-sensitive SFG spectroscopy allows us to extract information about molecular contact across different systems.

**Figures 5a,b** show the SFG spectra for the different PMMA substrates (**Table 1**) in contact with sapphire in both the hydrocarbon (C–H) regions at 2750–3100 cm⁻¹ and the hydroxyl (–OH) regions at 3100–3850 cm⁻¹. For comparison purposes, we also show the SFG spectra for the bare prism before contact ("Air"). Contact is made to pressures generally larger than those achieved in the ATR-IR experiments so as to allow the deformed contact spot to be fully within the beam (see the **Methods** section). In addition, a sapphire prism with a spuncast PMMA layer (where the solution is cast and annealed above the glass transition temperature, $T_g$) is provided as the most extreme case of molecular conformality. The fits to the raw spectra (solid lines) in both regions were accomplished using the Lorentzian function and the fitting parameters provided in **Table S3**. The C–H region shown in **Figure 5a** includes the aliphatic and aromatic vibrations from the chemical species present at the interface. The absence of any hydrocarbon signatures for the bare prism confirms a contaminant-free sapphire surface. Upon contact, distinct PMMA aliphatic signatures (*i.e.*, vibrations of ester-methyl groups and α-methyl groups) are observed. The characteristic peaks of PMMA observed across different samples include the symmetric vibration of the ester-methyl group at 2950 cm⁻¹ and the asymmetric vibration of the α-methyl group at 2995 cm⁻¹. The peak at 2960 cm⁻¹ corresponds to the asymmetric vibration of the backbone methylene signatures. Weak peaks in the 2850–2890 cm⁻¹ region correspond to the Fermi peaks of the ester-methyl groups(32). These signatures confirm the presence of PMMA at the contact interface.

Further, the fits show distinct differences in relative peak intensities of the characteristic peaks across different PMMA samples. The fitting parameters in **Table S3** show the presence of negative peaks of certain signatures across substrates. The negative amplitude for fitting parameters in some samples suggests the opposite orientation of the side groups at the interface.(33) Thus, all of these observations point toward the different local structures (conformation) adopted by the PMMA chains next to the surface. However, In **Figure 5a**, we can see that the relative aliphatic peak intensities are similar within the same group (spuncast, bilayers, and hardsheets). These differences in conformation would have direct implications on the molecules available for interaction at the surface, further affecting the molecular contact area.

The region from 3100–3850 cm$^{-1}$ in **Figure 5b** shows the O–H vibrations of the hydroxyl groups present on the sapphire interface. The air spectrum shows a sharp peak around 3710–3715 cm$^{-1}$ allocated to the free O–H vibration of the sapphire hydroxyl(26, 34). As illustrated in **Figure 5c**, on contact with PMMA substrates, these free hydroxyl vibrations shift to lower wavenumbers due to the acid–base interactions with the functional groups in PMMA(25, 26). The polar nature of the carbonyl group present in PMMA introduces a large shift (100–120 cm$^{-1}$) in the surface hydroxyl peak, leading to a broad-shifted hydroxyl vibration centered around 3600 cm$^{-1}$ (26). Thus, the increase of the shifted peak intensity with respect to the free O–H peak reflects on the number of new interactions introduced by contact, which suggests an increase in the molecular contact area between the PMMA molecules and the free hydroxyl present at the surface of the sapphire crystal.

To extract information about the molecular contact area, we define a parameter termed as *Shifted OH%* as the percentage of the total OH region comprised by the shifted OH signal. A detailed explanation of the calculation of this parameter is provided in the Supplementary Information. The *Shifted OH%* acts as a quantity related to the number of surface hydroxyl groups of the sapphire surface that are interacting with PMMA molecules at the molecular scale as compared to the total available number of hydroxyl groups within the region probed by the beams. In this way, *Shifted OH%* serves as an indication of the strength of the acid–base interaction between the two surfaces and, by extension, the percentage of the area between the two substrates that is in molecular contact. We add the caution that the calculated values, due to how the parameter is defined, are of a magnitude that is related to, but not directly equal to, the number of molecular contacts.

The calculated Shifted OH% values for each of the different types of contacts are shown in **Figure 6a**, where the values are referenced to the *Shifted OH%* of the same sapphire prism before contact was made. The results show decreasing values in the following order: Spuncast > Bilayers > Smooth Hardsheet > 30-grit Hardsheet. In the following discussion, we will try to understand the connection between these values and the molecular contact achieved for each system.

The spuncast film, representing the most extreme case in which we expect maximum possible molecular contact between PMMA and sapphire, shows the largest *Shifted OH%* as expected. In case of PMMA–PDMS bilayer mechanical contacts, we observe an intermediate *Shifted OH%,* indicating less interfacial interaction as compared to that for the spuncast film. However, we observe no pressure dependence of the *Shifted OH%* parameter (**Figure S8**) and note the fact that the two different bilayer thicknesses show indistinguishable values. This, alongside the lack of any discernable free hydroxyl peaks in the spectra for the two bilayer samples, suggests the plausibility that both bilayer samples achieve conformal molecular contact, despite the counterintuitive observation that they show lower interfacial interactions than the spuncast film. However, compared to mechanical contact, the process of the contact of the spin coating against sapphire allows for polymer chain rearrangement because the chains are placed in a solution state. Further, the annealing process ensures that a thermodynamically equilibrated state for the chains next to the interface is achieved(35). This equilibrated state allows for the greatest possible interactions, further maximizing the interfacial interactions with the sapphire hydroxyl groups beyond what can be achieved through conformal mechanical contact with PMMA below its $T_g$.

The smooth hardsheet, while smooth at macroscopic scales, has a much larger modulus. In **Figure 5b**, the presence of a sharp free hydroxyl peak suggests partial contact at the molecular scale. Further, the reduced molecular interaction due to the partial contact is reflected in the lower value of *Shifted OH%* as compared to that for the bilayer samples. The macroscopically rough 30-grit PMMA hardsheet, due to its greater roughness as compared to the smooth hardsheet, shows an even lower value and a visibly increased relative intensity of the sharp free hydroxyl peak— although the *Shifted OH%* value is still significant despite the expectation of a very minor molecular contact. Since the airgap thicknesses for the out-of-contact regions are large (**Figure 3a**), SFG intensity will be significantly reduced as compared to the in-contact regions due to differences in Fresnel factors at the chosen incident angle (**Figure S9**). As a result, the percentage of the shift is expected to be large even if minimal contact is achieved. Despite this, the shift is the smallest of all the samples.

To correct for Fresnel factors and quantitatively determine the percentage of molecular contact area achieved across each system, we employed a model described in the Supplementary Information to consider the *Shifted OH%* as a fractional sum of the value from the in-contact and out-of-contact sites weighted by any enhancements due to Fresnel factors. If the *Shifted OH%* from the 10-nm bilayer is taken as that for conformal mechanical contact, then the percentage of the area in contact can be calculated. The results of these calculations, shown in **Figure 6b**, indicate that the smooth hardsheet achieves around 45% ± 28% contact, while the 30-grit hardsheet achieves around 8% ± 13% contact. The value of 8% contact, though still large for this rough sample, is reasonable given the very large pressures that must be applied to achieve contact large enough for alignment with the incident beams. In addition, we further consider the fact that even the 10-nm bilayer sample is not able to achieve the maximum possible interfacial interactions with a *Shifted OH%* that is only 58% that of the spuncast film. As such, the contact area percentages are also considered as percentages of the maximum interaction strength achievable in **Figure 6c**.

**Discussion**

In this work, we have considered various samples consisting of PMMA with different effective moduli as thin layers on top of PDMS and hard, thick PMMA sheets of varying roughness. Through the application of imaging ATR-IR and SFG spectroscopy, we have characterized the nature of each contact at length scales from the micron level down to the subnanometer level, where one expects intermolecular interactions to occur.

For the high-modulus PMMA samples with varying roughness, we have shown that ATR-IR coupled with an imaging detector can resolve the heterogeneous nature of the contact, determine the compression of airgap separation with pressure, and estimate the real contact area percentage. We show an exponential drop in real contact area (**Figure 4d**) and average airgap separation (**Figure 4e**) with increasing RMS roughness. Furthermore, the smooth hardsheet sample represents a case where the small length scale of the roughness generates air gaps that are a small fraction of those for roughened surfaces (< 30 nm) and are instead dominated by the roughness of the germanium crystal, which becomes the rougher of the two surfaces. While real contact area percentages are calculated as 90%, such small-scale roughness could lead to an overestimation of real contact area when using a technique that is not surface-sensitive to these scales. However, from the SFG spectra (**Figure 5a,b**) and the analysis of the corresponding shifted OH% (**Figure 6b**), we show that a smaller value of the real contact area of around 45% ± 28% is achieved despite being obtained at a higher pressure. In addition, the 30-grit hardsheet shows a value of around 8% ± 13% contact compared to the value of approximately 6% ± 4% obtained from ATR-IR. As previously stated, the estimate from ATR-IR is expected to be more accurate for samples with larger roughness. Thus, it is logical that the results obtained from these two techniques should be comparable, especially given the larger pressure used in the SFG experiment.

The results for the smooth samples with varying effective modulus, presented in **Figure 3b**, show how the ATR-IR technique can allow us to distinguish between the different air gaps of the hard (hardsheet) and soft (bilayer) samples of different moduli due to their varying compliance. The technique is further able to resolve the 10-nm compression in air gap with pressure achieved for the smooth hardsheet as compared to the lack of any compression for the bilayer samples within the technique resolution. However, the results also highlight the airgap separation sensitivity, which is around 1–1.5 nm, and the absolute airgap resolution, which is around 10 nm. Since intermolecular interactions occur at scales smaller than 1 nm(4)·(36) and it is ultimately these interactions that govern the interfacial properties, SFG spectroscopy is especially useful tool for evaluating smooth samples, as it has a sensitivity specific to molecular-length scales that is achieved by probing the interactions directly. For example, while one might expect that a very thin layer of a stiff, glassy polymer (*i.e.*, a bilayer) should be able to bend and deform at an interface from a macro-scale perspective, we did not know how the locally high modulus might influence the conformability on molecular scales. With the help of the SFG spectra (**Figure 5a,b**) and the shifted OH% analysis (**Figure 6b**), we are able to show that the bilayers form a molecularly conformal contact. This finding is supported by the observation that two bilayer samples having different thickness of the stiff PMMA layer exhibit the same shifted OH% value and show no discernable free OH peak. In addition, the 200-nm bilayer does not show any changes in the shifted OH% values with pressure (**Figure S8**). From the results for the smooth hardsheet sample, we have also demonstrated that, even at the large pressures applied in our SFG experiments, the molecular contact that is made is only a fraction of that achieved for the bilayers.

Notably, the SFG results also reveal the importance of an additional parameter that is not often considered in discussions of real contact area: namely, the molecular conformation of the species in contact. For the sample with PMMA spuncast and annealed directly on the sapphire prism, an even greater shifted OH% value was measured that greatly exceeded the values obtained for either of the bilayer samples. While this finding might be presumed to be the result of the spuncast film conforming to a greater degree against the sapphire, this conclusion is not consistent with the pressure- and thickness-dependent bilayer data alongside the ATR-IR data for the bilayers, which suggest conformal molecular contact. Instead, the hydrocarbon regions of the SFG scans (**Figure 5a**) point toward different conformations of the PMMA chains at the interfaces produced by mechanical contact as compared to that for the spuncast interface. The conformation of the spuncast film, which maximizes the interactions across the PMMA–sapphire interface, leads to the higher value for the shifted OH%. Thus, the two systems that achieve complete coverage of sites at molecular length scales—the spuncast films and the bilayer contacts—have very different interaction energies that should, in turn, lead to differences in interfacial properties such as friction and adhesion. These intricacies are important to consider whenever the contact is achieved in different ways (*e.g.*, mechanical contact as opposed to solution/melt) or when a material is processed at or above its $T_g$ such that different degrees of molecular rearrangement can occur.

In summary, the composite knowledge gained from the ATR-IR and SFG characterizations paints a more complete picture of the nature of the contact interface down to molecular length scales for contacts ranging from soft and smooth to rigid and rough. Even though the actual contact area scales as a function of pressure as expected from the Bowden–Tabor description of friction, the actual contact area drops off exponentially with the increase in RMS roughness. This finding highlights the importance of the actual contact area in quantifying the friction coefficient and adhesion. The roughness and modulus have a much more dramatic effect on the actual contact area and the expected friction coefficient, even though the chemistry is identical for all of the rough and bilayer samples.

We further show results that highlight the distinction between contact of molecules with different orientations from mechanical contact versus solution state contact in terms of interfacial interaction strength. This knowledge, which could not be readily obtained using previously reported tools, is of critical importance to truly understand the means by which surfaces make contact. With this new understanding, techniques employing ATR-IR and SFG offer the ability to shed new light on fundamental processes ranging from friction and adhesion to haptics, nanofabrication, and solid-state batteries.

**Materials and Methods**

## Materials and Substrates

Poly(methyl methacrylate) (PMMA) with a molecular weight of 100,000 g/mol and a polydispersity of 1.09, obtained from Scientific Polymer Products (Ontario, NY, USA), were used for PMMA bilayer preparation. Additional PMMA with a molecular weight of 75,000 g/mol, obtained from Sigma-Aldrich (St. Louis, MO, USA), was used for thick PMMA sheet preparation. Sheets of polydimethylsiloxane (PDMS) were prepared according to previous reports[29]. Sandpapers of two grit sizes (30 grit and 1500 grit) were obtained from 3M (Maplewood, MN, USA). Additional chemicals of high purity were obtained from common commercial suppliers. All chemicals were used as received.

Equilateral sapphire prisms (15 mm × 15 mm × 15 mm × 10 mm; Meller Optics, Providence, RI, USA) used for SFG experiments were first cleaned by baking at 750 °C for 4 hours; subjecting them to ultrasonication in toluene, acetone, and ethanol for 1 hour each; and subsequently exposing them to plasma treatment for 5 min (PDC-001-HP expanded plasma cleaner; Harrick Plasma, Ithaca, NY, USA). For mechanical contact experiments, these prisms were directly used. To prepare spuncast PMMA prisms, we used a 4 wt% solution of PMMA in toluene and a rotation speed of 2000 rpm for 1 minute, followed by annealing at 120 °C overnight in vacuum.

PMMA–PDMS bilayers were prepared using the film transfer procedure previously reported[27]. The modulus of the chosen PDMS substrate was 1 MPa, and PMMA film (with thicknesses of either 10 nm or 200 nm) were transferred to the top of the substrate. Thick sheets of PMMA (PMMA "Hardsheets") were prepared by compression molding of stock PMMA powder between sheets of Kapton (smooth polyimide films) at 180 °C for 20 min under a load of 15,000 lbs to prepare a smooth surface. The final sheets, around 1 mm in thickness, were cut into pieces approximately 0.5 cm × 0.5 cm in size. For some samples, additional roughness was introduced by rubbing with a 30-grit or 1500-grit sandpaper sheet by hand with moderate pressure for 1 min. PMMA hardsheets were cleaned using ethanol and were blow-dried under $N_2$ flow prior to use.

## Roughness Characterization

The roughness values of PMMA hardsheets were measured using a Dextax KT stylus profilometer. Scans of 0.3 mm, 0.6 mm, and 1 mm were collected at multiple locations using a 2-μm tip. For soft bilayers, scans of 5 μm, 10 μm, 25 μm, and 50 μm were collected using a RTESPA 300 Tip (Bruker Corp., Billerica, MA, USA) of 8-nm nominal radius and 40-N/m stiffness using a Bruker Dimension Icon AFM. All the measurements for respective samples were stitched to obtain a power spectral density (**Figure S1**).

## Attenuated Total Reflectance Infrared Spectroscopy

Attenuated total reflectance infrared (ATR-IR) spectra were collected using a Cary 680 FTIR alongside a Cary 620 FTIR microscope accessory equipped with a focal plane array (FPA) detector from Agilent Technologies Inc. (Santa Clara, CA, USA). A germanium single hemispherical crystal with a radius of curvature of 3.2 mm was used for the contact experiments and served as the ATR crystal. An infrared beam incident at an angle of 29° to the germanium–sample interface was reflected after a single bounce into the FPA detector, which produces a 64 pixel × 64 pixel image with a spatial resolution of 1.1 μm × 1.1 μm per pixel, where each pixel captures the absorbance in the spectral range of 4000–900 $cm^{-1}$ with a resolution of 4 $cm^{-1}$. The germanium crystal before contact with the sample (in air) was used as the background scan. Contact pressure was controlled for these experiments in terms of the indentation depth of the germanium probe into the sample, which was controlled by a precision Z-stage and referenced to the initial point where contact was detected.

## Sum Frequency Generation Spectroscopy

Sum frequency generation (SFG) spectra were collected using a Spectra Physics laser system, which has been described in previous work[37]. Using this system, an 800-nm visible beam and an infrared beam tunable in the range of 2000–3850 cm$^{-1}$ with incident angles of 8.5° and 10°, respectively, were overlapped at the surface of a sapphire prism in total internal reflection geometry. The incident angles were chosen to provide a maximum signal contribution from the PMMA–sapphire contact interface based on Fresnel factor calculations. The spectra were collected in PPP polarization (i.e., P-polarized SFG, visible, and infrared beams). The prism was fitted into a custom contact cell that allowed for spectra to be collected with a sample brought in contact *in situ*. Experiments began by collecting a blank sapphire–air spectrum for reference followed by bringing the substrate of interest into contact with the sapphire using a stepper motor with a net displacement of around 2–4 μm to create a contact spot that is larger than the size of the laser beam. The saturation of increasing SFG signal with pressure provided confirmation that the contact spot was sufficiently large. We estimate that the contact pressure achieved for the hardsheet samples was around 50 MPa and the contact pressure achieved for the bilayer samples was around 20 kPa. Two scans for each sample at a given condition were averaged, and three repeats of each substrate in contact were collected.

**Acknowledgments**

The authors acknowledge financial support for this work from the National Science Foundation (Award No. DMR-2208464). AD also acknowledges financial support from the Knight Foundation (W. Gerald Austen Endowed Chair). The authors also thank E. Laughlin for custom device fabrication, P. Karanjkar for help with AFM, Chaitanya Gupta for help with the compression molding of the PMMA sheets, G. Mallinos and V. Naiker for help with SFG, and the Avery Dennison Corporation for graciously donating the FTIR microscope and the FPA detector used in this study.

**Figures and Tables**

**Table 1:** Modulus and root mean square roughness data for different substrates. The elastic modulus for bilayer samples is calculated using rule of mixture(38)·(39). The root mean square roughness for different PMMA substrates were measured using stylus profilometry ($^*$); for bilayers, AFM ($^\dagger$) was used for this purpose.

| Sample | Modulus | $h_{rms}$ (Roughness) |
|---|---|---|
| 10-nm Bilayer | 1.0 MPa | 2.5 ± 0.5 nm$^\dagger$ |
| 200-nm Bilayer | 1.6 MPa | |
| Smooth Hardsheet | 2.9 GPa | 5.9 ± 3.8 nm$^*$ |
| 1500-grit Hardsheet | 2.9 GPa | 290 ± 13 nm$^*$ |
| 30-grit Hardsheet | 2.9 GPa | 1400 ± 800 nm$^*$ |

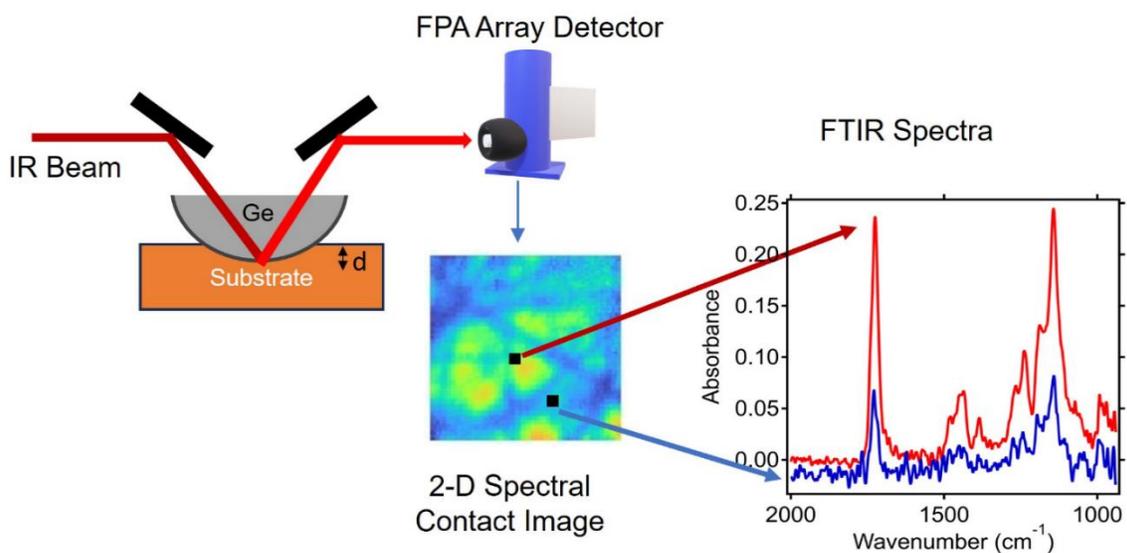

**Figure 1.** Diagram of the ATR-IR imaging experiment, alongside an example of a two-dimensional (2D) spectral image collected for germanium in contact with a 1500-grit PMMA hardsheet. Here, each pixel in the resulting 64-pixel × 64-pixel image corresponds to an independent ATR-IR absorbance spectrum—two of which are provided as examples on the right. The example image is constructed based on differences in intensity for the 1725 cm$^{-1}$ peak assigned to PMMA. The dimensions of the image are 70 μm × 70 μm.

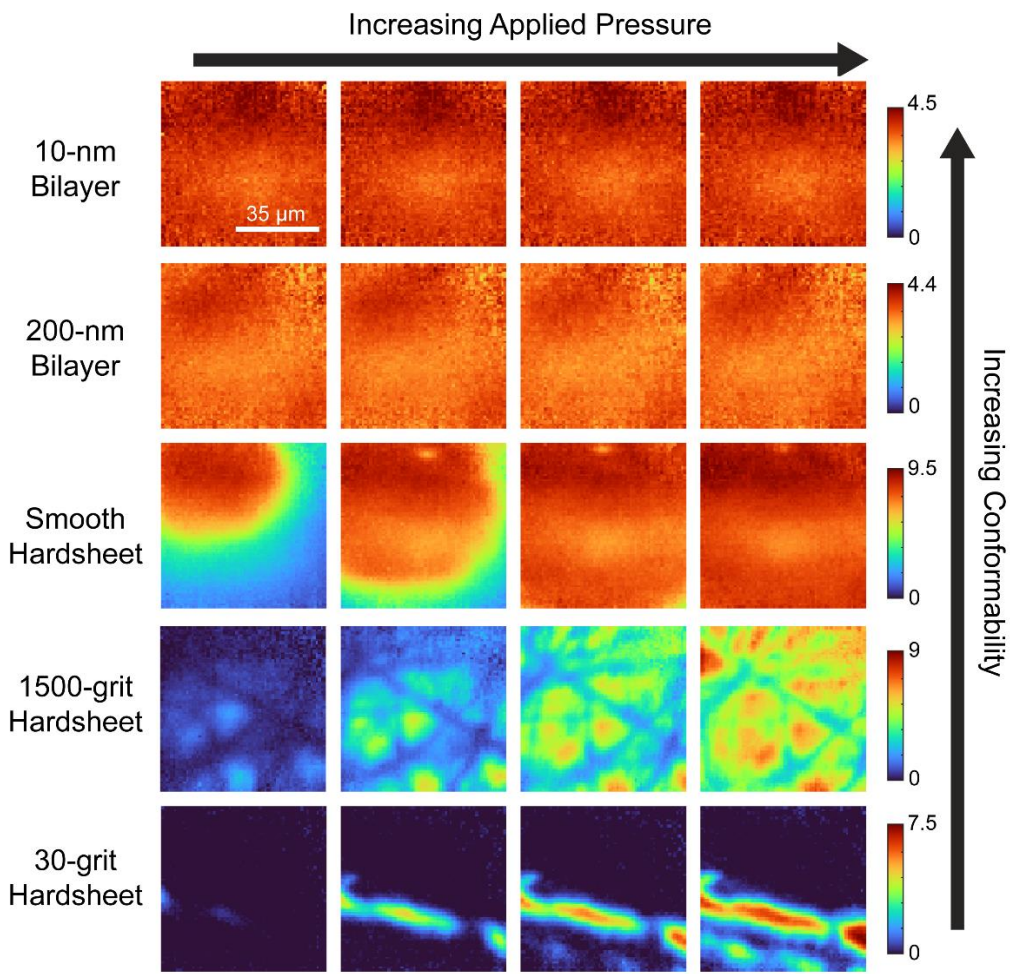

**Figure 2.** Representative images of the PMMA–germanium contact for different sample types of varying modulus and roughness and at different pressures. The samples are arranged with increasing conformability from bottom to top, while the pressures are arranged with increasing pressure from left to right. Each image is constructed using the calculated area for the 1725 cm$^{-1}$ peak assigned to PMMA, apart from the 10-nm bilayer, where the peak at 1260 cm$^{-1}$ assigned to PDMS is used. The peak area color scale is provided for each sample type. The different scales for each sample type are indicative of either the different PMMA layer thicknesses, the different peaks analyzed, or the roughness-induced air gaps. Direct comparisons of peak areas only apply within a given sample type.

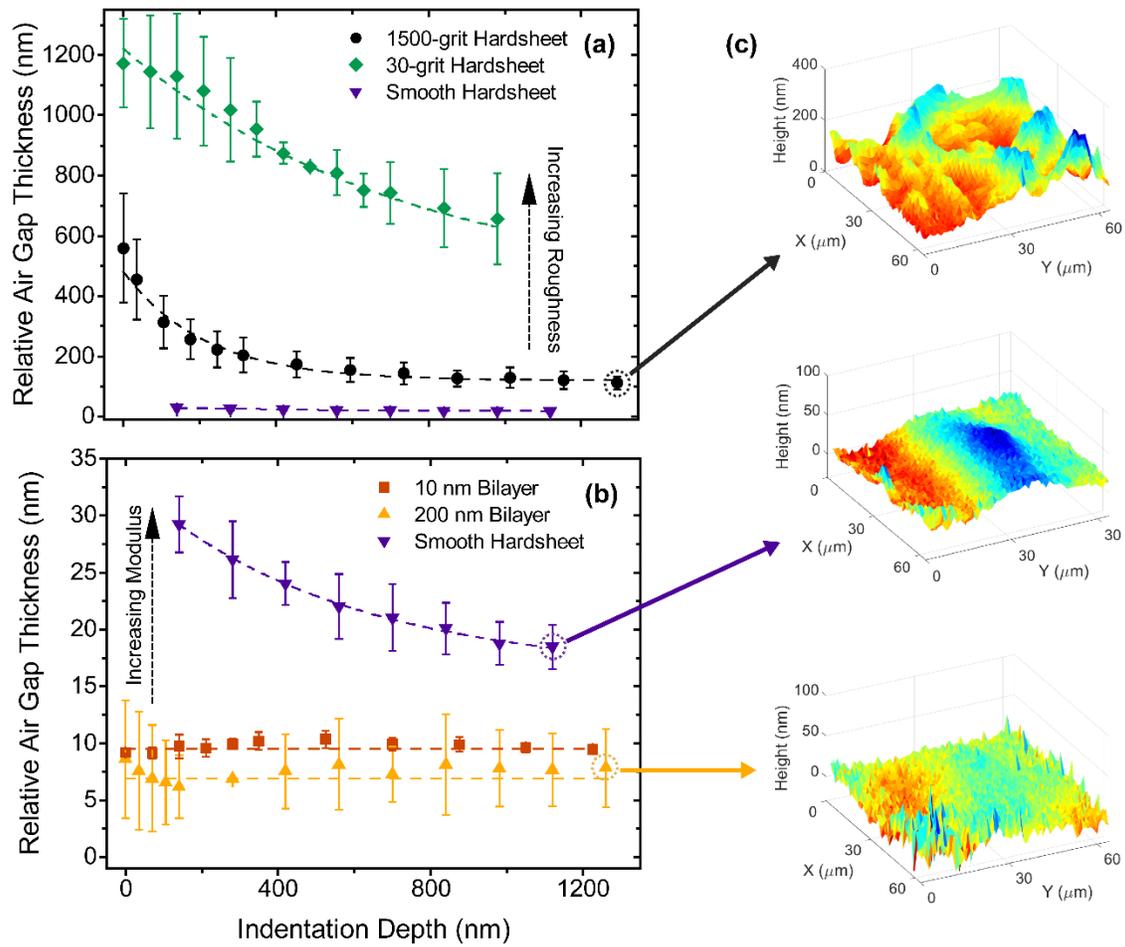

**Figure 3.** Calculation of average area gap thickness within the analysis region vs. the applied indentation depth of the germanium crystal into the sample for **(a)** two sandpaper–roughened PMMA hardsheets (30-grit and 1500-grit) and **(b)** smooth hardsheet and PDMS–PMMA bilayers. **(c)** Three-dimensional height maps of the air gaps, which provide spatial information of the contact interface for the maximum pressure of 1500-grit hardsheets, the smooth hardsheets, and the 200-nm bilayer samples.

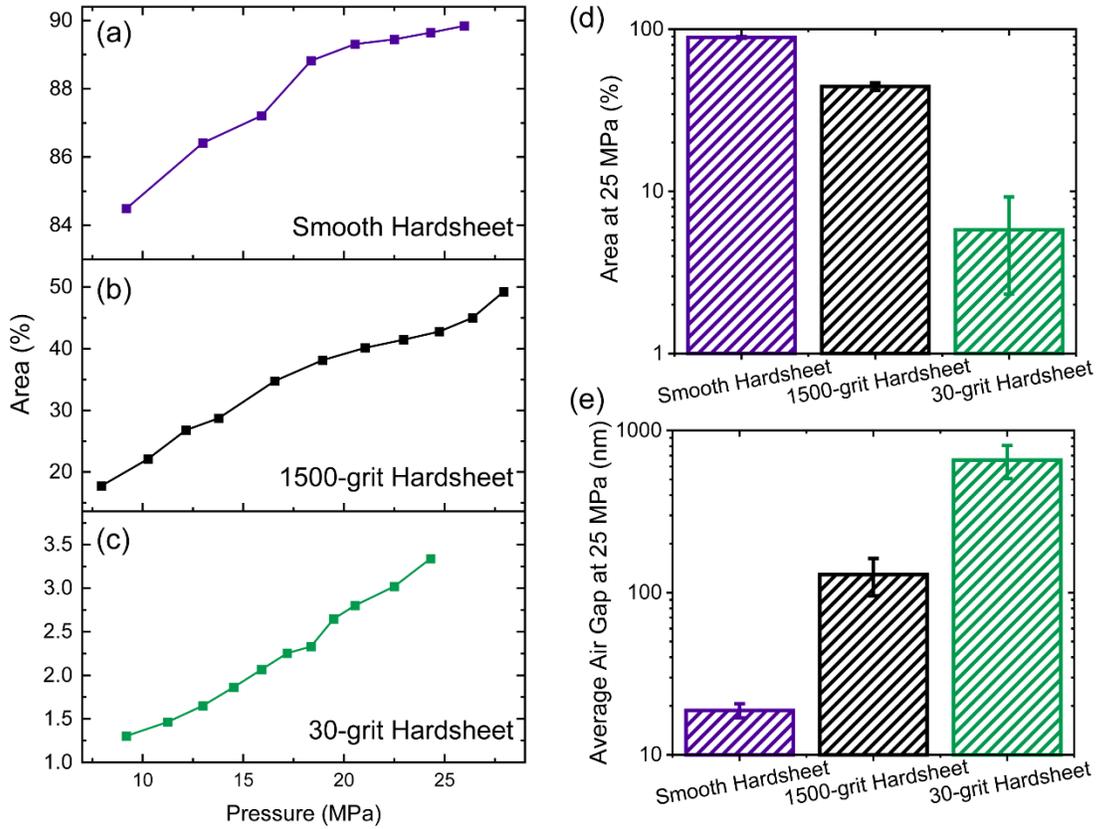

**Figure 4.** Real contact area analysis results as contact area percentage (the percentage of the total frame area where surfaces are in contact) as a function of applied pressure for samples of different roughness: **(a)** smooth hardsheet, **(b)** 1500-grit hardsheet, and **(c)** 30-grit hardsheet. **(d)** Log-scale chart comparing the contact area percentage values between samples of different roughness at an applied pressure of 25 MPa. **(e)** Log-scale chart comparing average airgap separation between samples of different roughness at an applied pressure of 25 MPa.

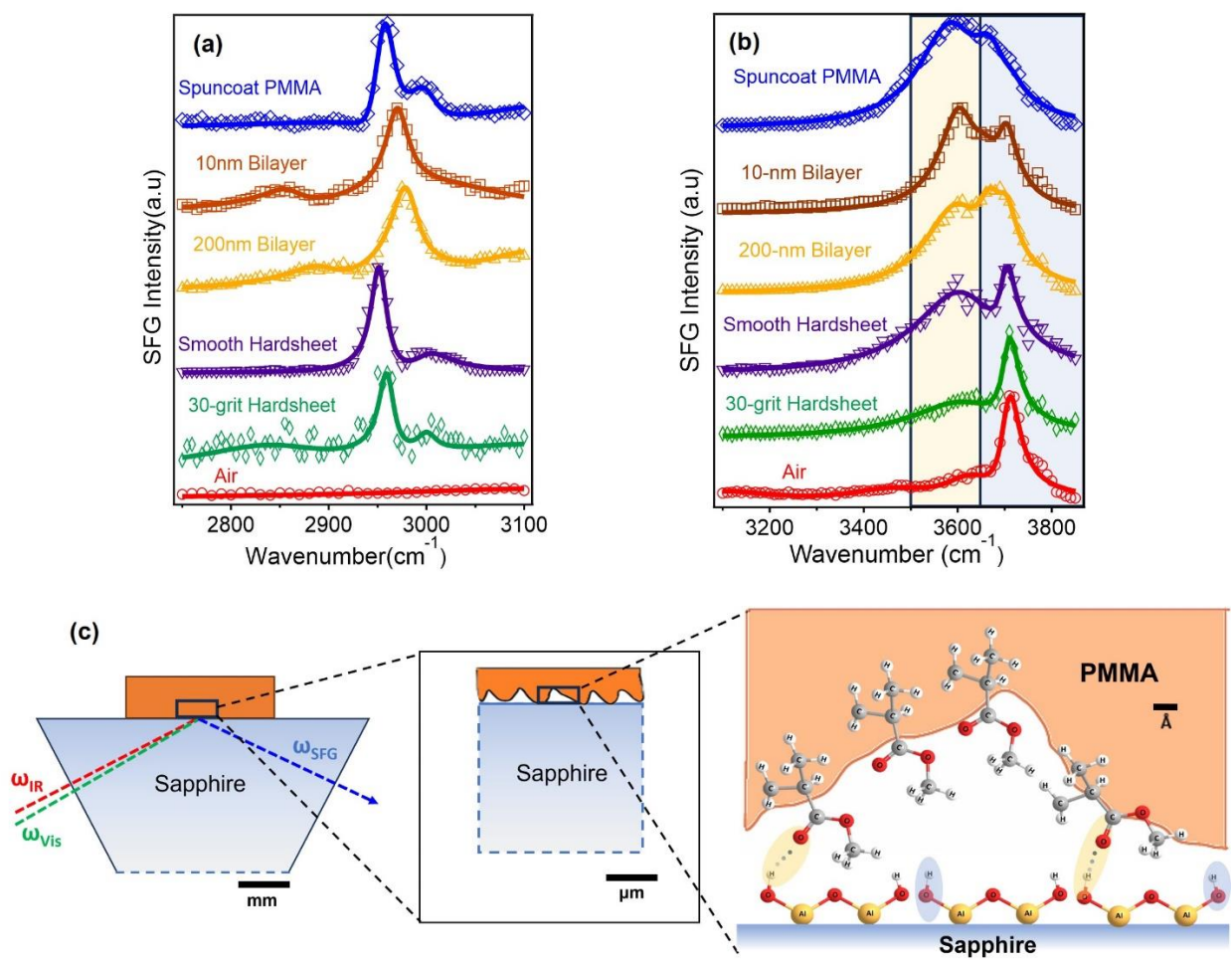

**Figure 5.** SFG spectroscopy for understanding the molecular contact between different PMMA substrates. The top portion of the figure shows the raw spectra and fits to the data in **(a)** the C–H region (2750–3100 cm$^{-1}$) and **(b)** the –OH region (3100–3850 cm$^{-1}$). The shaded regions in (b) show the regions used for further analysis including the water region (*white*), the shifted OH– region (*yellow*) and the free –OH region (*blue*). **(c)** Schematic diagram showing the SFG setup, wherein IR ($\omega_{IR}$) and visible ($\omega_{Vis}$) beams are incident on the contact region. The SFG beam that originating at the surface ($\omega_{SFG}$) was collected to give the raw SFG spectra. The schematic illustrates the nature of molecular contact with the carbonyl groups (*yellow*) of PMMA as they interact with the surface hydroxyl groups (*blue*) on the sapphire substrate. These acid–base interactions are short-range interactions and are direct indicators for molecular contact.

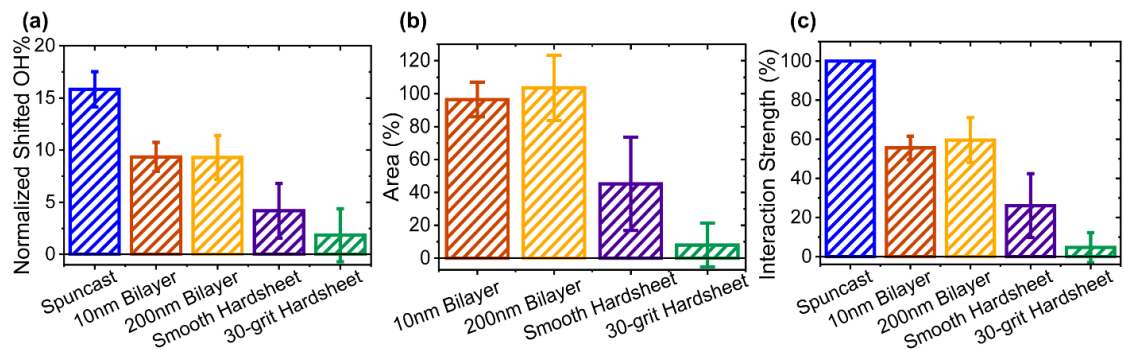

**Figure 6.** Variation of **(a)** *Normalized Shifted OH%* parameter across different PMMA-based surfaces, **(b)** real contact area as a percentage of the nominal area calculated using the *Shifted OH%*, and **(c)** interaction strength percentage as the percentage of the maximum possible interaction achieved at the contact interface for each system.

**Supporting Information for**

Measurement of Nanoscale Interfacial Contact Area Using Vibrational
Spectroscopy

Utkarsh Patil[1†], Stephen Merriman[1†], Shubhendu Kumar[1†], Ali Dhinojwala[1*]

*Corresponding author, **Email:** ali4@uakron.edu

†These authors contributed equally to this work.

**This PDF file includes:**

Supporting text

Figures S1 to S9

Tables S1 to S3

SI References

**Supporting Information Text**

**Calculation of air-gap thickness from ATR-IR data.**

To further analyze differences between the contacts with different samples and as a function of applied pressure, peak area was converted to airgap thickness. We calculated the dependence of absorbance peak area on air gap thickness between the germanium and sample using a multi-layer optical model. The details of this model, along with the optical constants used for each material layer, are provided in Table S2. Using this method directly, the calculated values for peak area at zero air gap thickness were generally not comparable to those measured experimentally. Reasons for this include the fact that the IR beam incident angle is not precisely known by the nature of the instrument design from the manufacturer of the ATR microscope, and the fact that the optical constant values taken may differ slightly from those of our actual materials. However, the shape of the curve does not depend strongly on the choice of incident angle and optical constants (Figure S4). As a result, the calculated curves were scaled assuming that the peak area at zero airgap corresponded to the measured value for the most intense region of the image for each sample type at the maximum applied pressure. For the hardsheets, this assumption was only made for the smooth sample (*i.e.*, we assumed the peak area of zero airgap for the smooth hardsheets applied to the rough hardsheets). This region is shown in Figure S3 and was kept fixed across samples. The region was chosen as each of the smooth samples showed, on average, the highest peak areas in this region due to its far proximity from the surface asperity on the germanium as discussed in the main text. The validity of this assumption is further discussed below from the calculated results. The final curves representing the relationship between peak area and air gap thickness are presented in Figure S5. Using these curves, the peak areas for each pixel were converted to airgap thickness, with the average values for the whole frame shown in Figure 3a,b and the pixel distributions provided in Figure 3c. Importantly, since the apparent area of contact (*e.g.*, the circular contact spot seen for the smooth hardsheet in Figure 2) does not always fill the whole frame at lower pressures, images where this apparent contact did not fill the entire frame were discarded from the analysis shown in Figure 3a,b. Thus, the analysis considers only airgap changes occurring within the apparent contact.

Finally, we consider whether the assumption of zero airgap being achieved for the bilayer and smooth hardsheet samples at maximum pressure is justified for the calculated values presented in Figure 3. In the case of the bilayer samples, the fact that we observe no intensity change (and thus no airgap thickness change) with pressure itself implies no airgap within our degree of detection exists. This observation is consistent with the data collected from SFG which also suggested no change in molecular contact with pressure and thus conformal contact for these samples. For the smooth hardsheet, we do observe a small change in intensity with pressure. While this change appears to plateau, this is no guarantee that the airgap achieved is actually zero even in the most intense peak area region which is further verified by the SFG results which indicate a significant lack of molecular contact. Although, this does suggest that any remaining airgap in the lowest airgap region should be small due to the small magnitude intensity changes and is indeed reasonable based on the small values of $h_{rms}$ (Table 1) and roughness PSDs (Figure S1). If a small airgap still exists at a 1200 nm indentation (30 MPa) in this intense region, the reported curve remains valid but with a small vertical shift. Thus, the values are reported as relative airgap thicknesses. Despite this, the reported values are not expected to be significantly different from true values, within the discussed precision of the technique. Further, such a small vertical shift would be insignificant for the roughened hardsheets due to the scale of the airgap thickness values.

**Calculation of Normalized Shifted OH% from SFG data.**

The basis of our analysis of OH peak shift from SFG spectra derives from the shift in the spectral signature of the OH group on the molecular surface of sapphire which occurs due to acid-base intermolecular interactions, which has been established in prior works(1–3). In our discussion of sapphire OH peak shifts, we refer to the presence of both *free OH* vibrations centered around 3710 cm$^{-1}$ due to non-interacting OH groups, and *shifted OH* vibrations centered around 3600 cm$^{-1}$ for the case of PMMA due to interacting OH groups. The width of the shifted peaks describes the distribution of possible interactions of the molecules; likewise, the sharpness of the free O-H vibration describes the bound nature of the surface hydroxyl groups on sapphire. Similarly, the multimodal, broad nature of the shifted OH peak centered around 3600 cm$^{-1}$ is attributed to the different intensity of the interactions (strong/weak) between the free-hydroxyls on sapphire with the carbonyl of PMMA. However, it does not account for the exact number of carbonyl groups in direct contact(4). Thus, the use of multiple fitted shifted OH peaks for analysis would overestimate the number of carbonyl peaks in direct contact with surface hydroxyl. We therefore use the concept of area under the curve for gauging the changes between the strongly shifted hydroxyl groups (centered around 3600 cm$^{-1}$) and the surface hydroxyl of sapphire (around 3710 cm$^{-1}$) after contact. We additionally consider the fact that absorbed water may interact with the sapphire free hydroxyl peak by taking the region centered near 3300-3400 cm$^{-1}$ to describe the presence of water based on previous reports(5)**.** These areas would point towards the proportions of total available surface hydroxyl divided into strongly polar interacting molecular contact, water molecules present due to humid surrounding environments, and free hydroxyl vibrations due to absence of molecular contact.

To extract the quantitative information of these areas, the hydroxyl spectral region is subdivided into three different regions depending on the presence of different molecular vibrations of the surface hydroxyl shown by the color shaded regions in Figure 5b. The three sub-regions are named as the *water-region* (3100-3500 cm$^{-1}$, white), the *shifted OH-region* (3500-3650 cm$^{-1}$, yellow) and the *free OH-region* (3650-3850 cm$^{-1}$, blue). The demarcation of these sub-regions is mentioned Figure S8 after considering the presence of free hydroxyl, shifted hydroxyl and water peaks observed in the SFG spectra across samples. These bounds are user-defined for extracting maximum contribution from the designated groups without introducing external bias, but they do not represent actual boundaries in the spectral sense. Area in each region is determined from the square root of SFG intensity vs. wavenumber, since it is the square root of SFG intensity that trends with concentration based on the fitting equation (Table S3).

We define a parameter called *Shifted OH%* for comparing the relative differences across the different substrates. This parameter is calculated in equation S1 wherein the Area $A_{shifted}$ corresponds to the area under the curve of the square root of SFG intensity vs. wavenumber in the Shifted OH-region, while the area $A_{total}$ corresponds to the total area under the curve of the OH region (sum of all three regions).

$$Shifted\ OH\% = \frac{A_{shifted}}{A_{total}} \tag{S1}$$

In Figure 5b, we observe that the air spectrum has non-zero area under curve in the shifted OH (3600 cm$^{-1}$) and water regions (3400 cm$^{-1}$) in the air spectrum. This suggests the presence of adsorbed water due to a humid environment, which may in general show signatures in both regions(6). Thus, the non-zero *Shifted OH%* in the reference air spectra which can generally vary between samples or under different ambient humidity necessitates the normalization of the *Shifted OH%* parameter for each individual sample using equation S2 to better compare between samples. The underlying assumption is that the same amount of adsorbed water is present before and after contact which should be justified for a highly hydrophilic surface such as sapphire.

$$Normalized\ Shifted\ OH\% = Shifted\ OH\%\ (Sample) - Shifted\ OH\%\ (Air) \qquad \text{(S2)}$$

**Calculation of Area (%) and Interaction Strength (%) from SFG data.**

The *Shifted OH%* values calculated from equation S1 can be used to determine the real contact area as a percentage of the nominal area by treating the *Shifted OH%* as a superposition of contributions from in contact and out of contact sites. Equation S3 describes this generalized treatment where $\chi$ is the fraction of sites in contact while *Shifted OH% (Contact)* and *Shifted OH% (Air)* are the *Shifted OH%* values for the in contact and out of contact (*i.e.*, air) sites, respectively.

$$\begin{aligned} Shifted\ OH\%\ (measured) \\ = \chi \times Shifted\ OH\%\ (contact) + (1 - \chi) \times Shifted\ OH\%\ (Air) \end{aligned} \qquad \text{(S3)}$$

Since air spectra are recorded for every sapphire prism before bringing any sample into contact, the *Shifted OH% (Air)* for each sample is known. Furthermore, since the results seem to indicate that the bilayer samples are mechanically conformal against the sapphire within the range of pressures considered in our experiments, it is reasonable to take the *Shifted OH%* value measured for the bilayers as the value for *Shifted OH% (contact)*. The value of $\chi$ can then be calculated for any value of *Shifted OH% (measured)*. We note that *Shifted OH% (Measured) – Shifted OH% (Air)* is the same as *Normalized Shifted OH%* from equation S2.

However, equation S3 assumes that an in contact site and an out of contact site provide equal signal contribution to the overall observed SFG spectrum. Although, as shown in Figure S9, signal is expected to decay rapidly as a function of air gap thickness between the sapphire and PMMA at the chosen incident angle for our experiments. For the smooth hardsheet, the h$_{rms}$ value of 2.5 nm (Table 1) and air gap thicknesses of 20 nm recorded from ATR-IR at the larger pressure of 25 MPa (Figure 4e) imply that the signal contribution between in contact and out of contact sites should be negligibly different since the scale of the out of contact airgaps will be very small. For the 30-grit hardsheet, the h$_{rms}$ value of 1400 nm and air gap thickness of ~700 nm at 25 MPa recorded from ATR-IR indicate that out of contact airgaps will create a substantial signal contribution difference. Since the trend in Figure S9 reaches a bottom plateau past 700 nm, we can approximate that there will be a 2 times relative signal enhancement for in contact sites (0 nm airgap) compared to out of contact sites (> 700 nm airgap). This leads to the adjusted form of equation S3 given in equation S4 where $f$ is the relative signal enhancement for in contact sites (~1 for smooth hardsheet and bilayers, ~2 for 30-grit hardsheet). The denominator of equation S4 is a normalization factor as the sum of the coefficients before each *Shifted OH%* term to ensure the addition of $f$ changes only the relative contribution and not the overall magnitude. Equation S4 is used to calculate the Area (%) values in Figure 6b where Area (%) is $\chi * 100\%$.

$$Shifted\ OH\%\ (measured)$$
$$= \frac{f\chi \times Shifted\ OH\%\ (contact) + (1 - \chi) \times Shifted\ OH\%\ (Air)}{f\chi + 1 - \chi} \qquad (S4)$$

Lastly, Area (%) values were converted to interaction strength (%) values. Since the *Normalized Shifted OH%* value from Figure 6a of the spuncast sample (maximum interaction strength overall) is 30.2% ± 4.3% compared to 17.4% ± 1.3% for the bilayers (maximum interaction strength for mechanical contact), this means that complete mechanical contact achieves only 57.6% of the interaction achieved by the spuncast sample since $f = 1$ for both samples. Therefore, the interaction strength (%) values for each of the mechanical contacts should simply be 57.6% × Area (%), which corresponds to the values reported in Figure 6c.

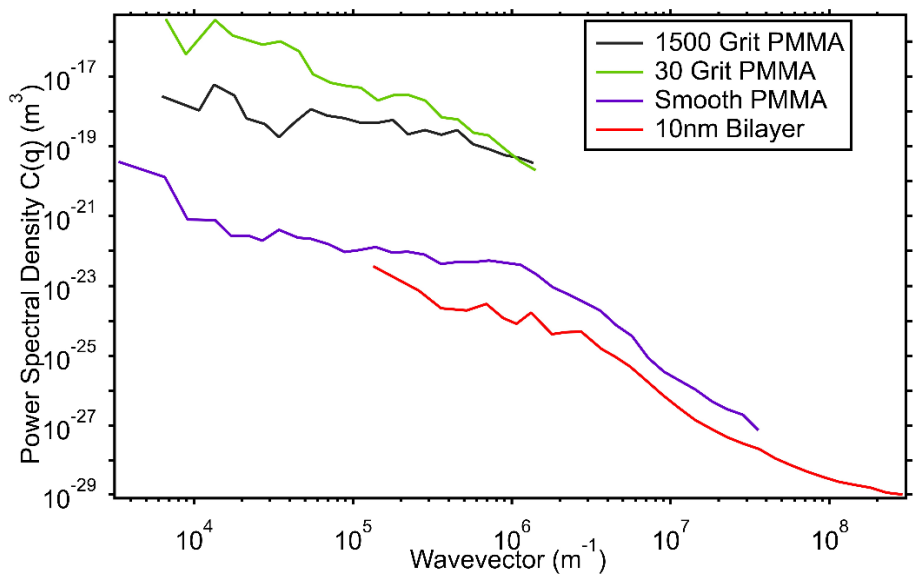

**Figure S1:** Power Spectra Density vs. Wavevector data for different substrates. Stylus Profilometry was used to measure roughness profiles for PMMA hard substrates while AFM was used for the soft bilayers. All the measurements were stitched to obtain a power spectral density using the web-based application contact.engineering(7, 8).

**Table S1:** Baseline points and integration bounds used to determine the peak area used for analysis of ATR-FTIR data. Baseline points are defined as a region, and the average value within the given region is used to define the position of the baseline point for each spectrum.

| Peak Assignment | Left Baseline (cm$^{-1}$) | Right Baseline (cm$^{-1}$) | Left Integration Bound (cm$^{-1}$) | Right Integration Bound (cm$^{-1}$) |
|---|---|---|---|---|
| PMMA CO | 1520-1650 | 1800-2000 | 1675 | 1775 |
| PDMS Si-CH$_3$ | 1190-1220 | 1310-1400 | 1230 | 1280 |

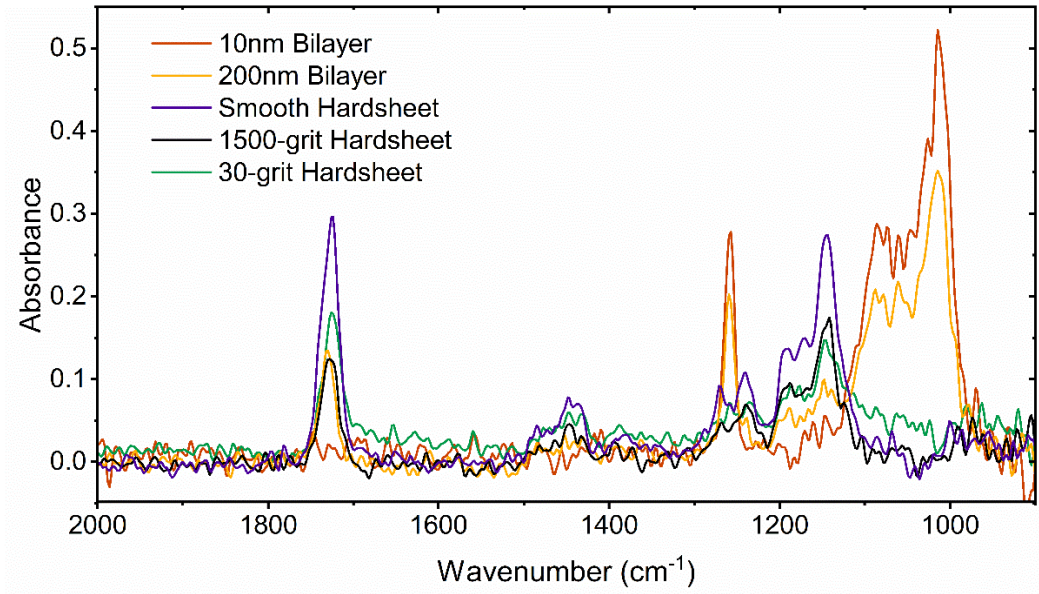

**Figure S2:** Examples of ATR-FTIR spectra within the region 2000-900 cm⁻¹ from FPA detector images collected for each sample type at the maximum indentation depth considered (~ 1200 nm).

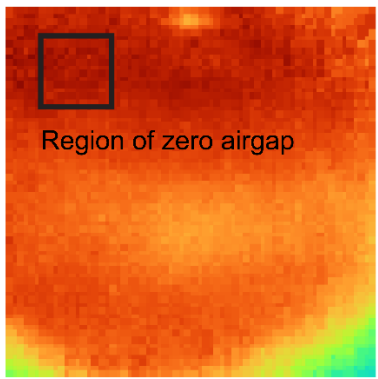

**Figure S3:** Images demonstrating the region (black box) used for determining the peak area corresponding to zero airgap thickness. The region was chosen as the most intense region consistent from sample to sample across the smooth samples and kept constant. Image dimensions are 70 μm x 70 μm.

**Table S2:** Optical multi-layer model and optical constants used for air gap calculations.

| Sample Type | Layer 1: Germanium | Layer 2: Air | Layer 3: PMMA | Layer 4: PDMS |
|---|---|---|---|---|
| 10 nm Bilayer | ∞   nm | 0-1000 nm | 10 nm | ∞   nm |
| 200 nm Bilayer | ∞   nm | 0-1000 nm | 200 nm | ∞   nm |
| Hardsheets (Rough and Smooth) | ∞   nm | 0-1000 nm | ∞   nm | 0 nm |
| **Optical Constants** | Amotchkina et al.(9) | n=1, k=0. | Zhang et al.(10) | Zhang et al.(10) |

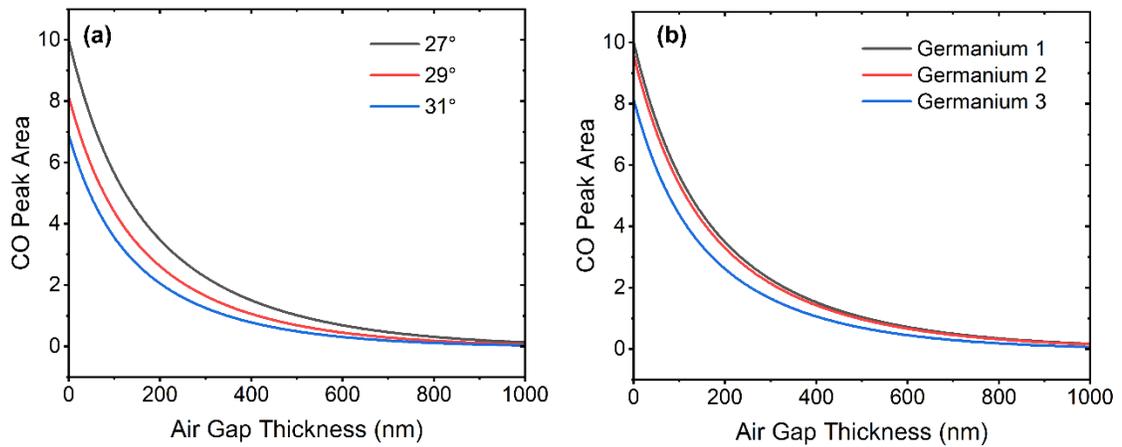

**Figure S4:** Raw air gap calculation results of PMMA CO peak height vs. air gap thickness using hardsheet layer parameters (Table S3) and different calculation parameters. **(a)** shows calculation results varying the angle of incidence for the IR beam. **(b)** shows calculation results varying the source used to obtain the optical constants of the germanium layer (Germanium 1(9), Germanium 2(11), and Germanium 3(12)). While the magnitude of the peak area changes considerably, the shape of the curve remains largely unchanged.

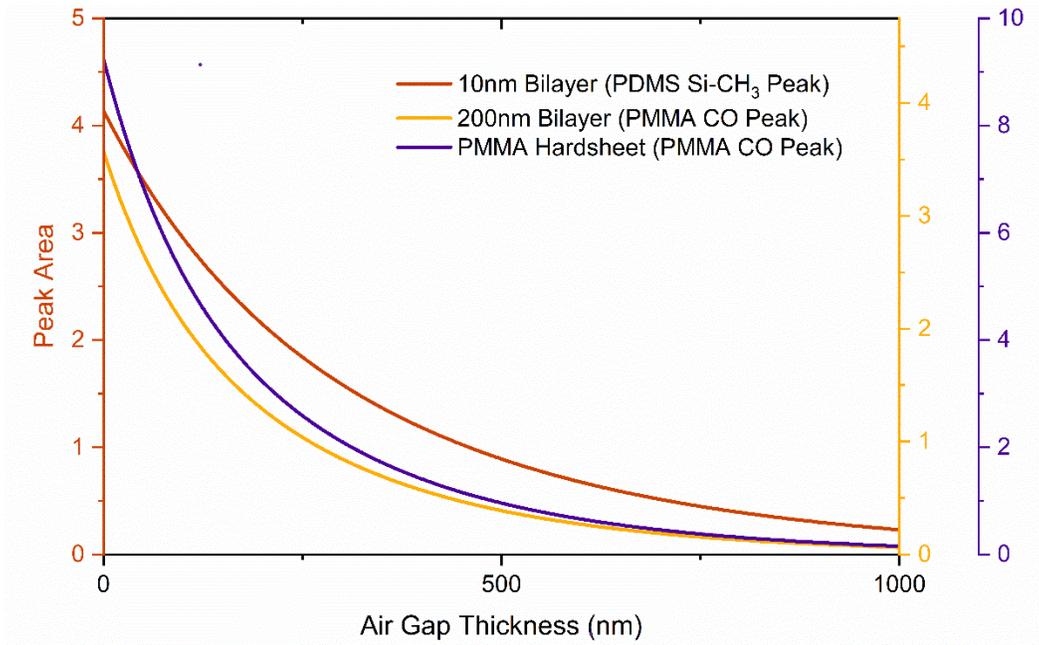

**Figure S5:** Peak area vs. air gap thickness from the multi-layer absorbance optical calculations for each sample type. For the 10nm bilayer, the Si-CH$_3$ absorbance peak assigned to PDMS is used. For the 200nm bilayer and PMMA hardsheets, the CO absorbance peak assigned to PMMA is used.

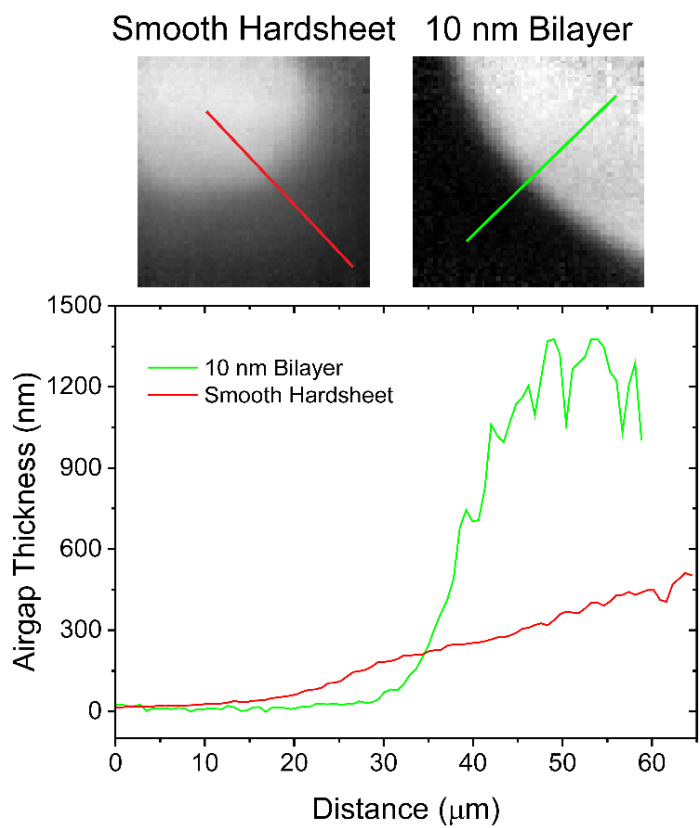

**Figure S6:** Monochrome images where intensity corresponds to absorbance peak area for smooth hardsheet and 10 nm bilayer for instances where the edge of the apparent contact is observed (Top). To explore the airgap profile across and away from the contact boundary, the intensity within the lines (red for smooth hardsheet, green for 10 nm bilayer) indicated in the top images was converted to airgap thickness values and plotted as a function of distance along the line (Bottom). The rapidly increasing airgap for the bilayer is consistent with a JKR adhesive-type contact, whereas the gradually increasing airgap for the smooth hardsheet is consistent with a hertzian, non-adhesive, hard contact.

**Table S3:** SFG fitting parameters from Lorentzian fits of CH region spectra in Figure 5a.

$$I_{SFG} \propto \left| \chi_{NR} + \sum_{q=1}^{Q} \frac{A_q}{\omega_{IR} - \omega_q + i\Gamma_q} \right|^2$$

| Fitting parameter | Spuncoat | 10nm Bilayer | 200nm Bilayer | Smooth Hardsheet | 30-grit Hardsheet |
|---|---|---|---|---|---|
| $\omega_1$ | 2949.96 | 2875.00 | 2957.54 | 2894.49 | 2850.00 |
| $\omega_2$ | 2960.00 | 2960.00 | 2965.00 | 2952.00 | 2960.00 |
| $\omega_3$ | 2999.00 | 3000.00 | 2995.00 | 3000.00 | 3000.00 |
| $\omega_4$ | 3150.00 | 3150.00 | 3100.00 | 3100.00 | 3100.00 |
| $\Gamma_1$ | 16.07 | 66.15 | 33.21 | 29.42 | 75.02 |
| $\Gamma_2$ | 15.36 | 17.90 | 40.90 | 10.00 | 9.56 |
| $\Gamma_3$ | 17.35 | 35.00 | 44.36 | 25.00 | 13.00 |
| $\Gamma_4$ | 140.00 | 110.00 | 110.00 | 78.72 | 218.95 |
| $A_1$ | -201.64 | -686.38 | -1541.08 | 0.00 | 74.61 |
| $A_2$ | 210.72 | 599.88 | 2364.48 | 349.36 | 21.73 |
| $A_3$ | 71.36 | -267.42 | -383.01 | 304.38 | 8.36 |
| $A_4$ | 900.00 | -1506.87 | 958.82 | 0.00 | 519.17 |

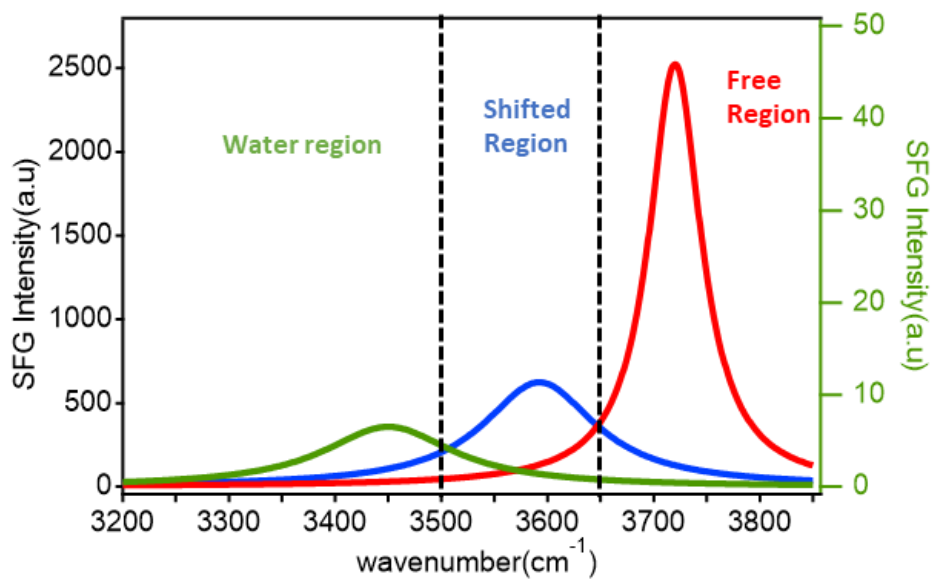

**Figure S7:** Demarcation between different regions based on the observed peak widths. The regions are defined by the intersection of the curves in green (3400 broad peak of loosely bound water), blue (broad shifted hydroxyl peak due to molecular interactions) and red (sharp 3700 free hydroxyl peak). The intersections around the wavenumbers take place roughly around 3500 cm[-1] and 3650 cm[-1] that serve as the bounds for the analysis of Shifted OH% in the main text.

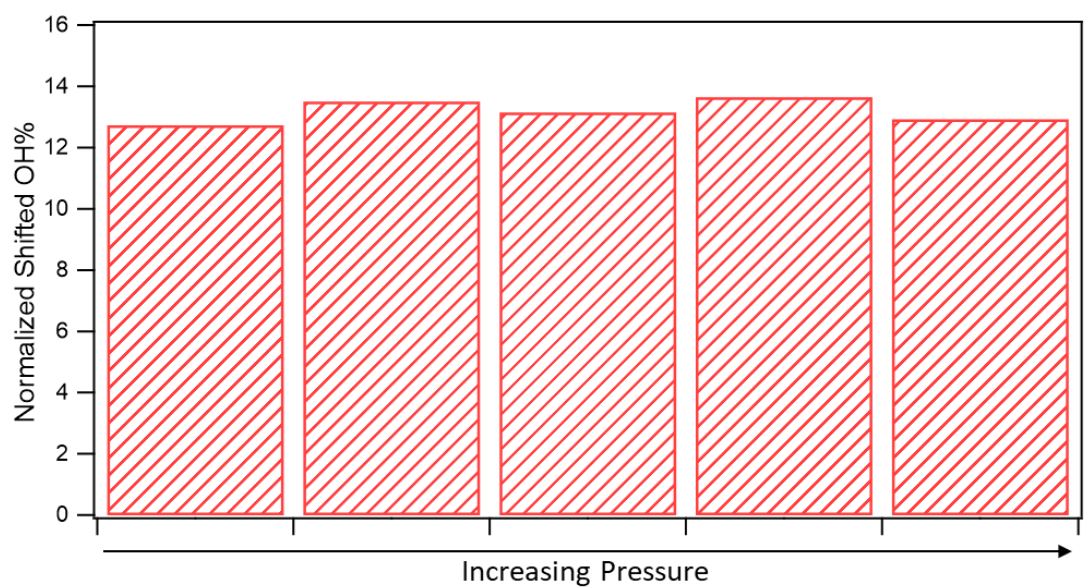

**Figure S8:** Plot of Normalized Shifted OH% vs pressure dependence for the 200nm Bilayer. The x-axis shows increasing pressure from P0 to Pmax with no significant differences in the Shifted OH%. This hints the complete possible molecular contact for 200nm Bilayer even with no pressure dependence.

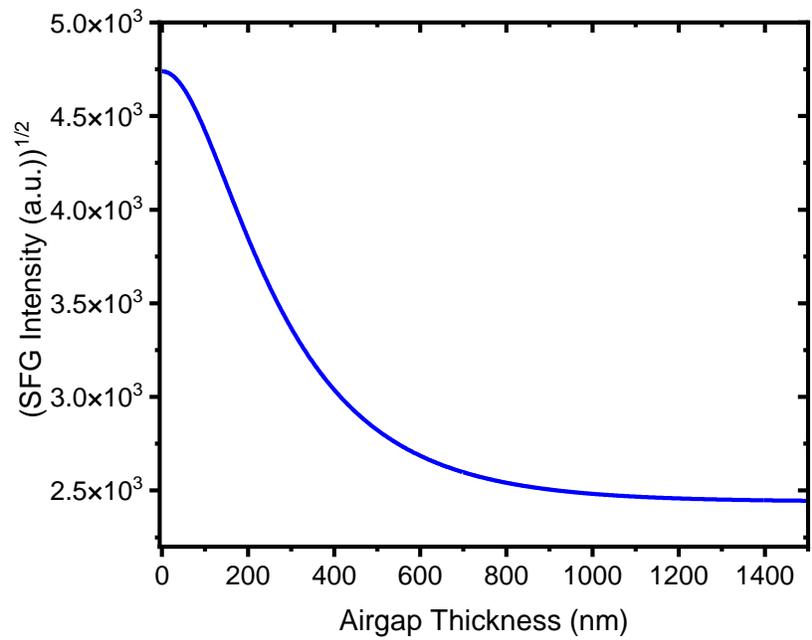

**Figure S9:** Calculated square root SFG intensity as a function of airgap thickness for a layer of air separating sapphire and PMMA at the incident angle for the reported experiments.

**SI References**